\RequirePackage{fix-cm}
\documentclass[twocolumn]{svjour3}          
\smartqed  
\usepackage{graphicx}
\usepackage{upgreek}
\usepackage{enumitem}
\usepackage{color}
\usepackage[normalem]{ulem}
\usepackage{amsmath}
\begin{document}

\title{Construction of Continuous Magnetic Cooling Apparatus with Zinc Soldered PrNi$_5$ Nuclear Stages
}

\titlerunning{Construction of Continuous Magnetic Cooling Apparatus}        

\author{S. Takimoto\and
        R. Toda\and
        S. Murakawa\and
        Hiroshi Fukuyama 
}


\institute{S. Takimoto \at
              \email{takimoto@crc.u-tokyo.ac.jp} 
              \and
              H. Fukuyama \at
              \email{hiroshi@kelvin.phys.s.u-tokyo.ac.jp} 
              \and
              Cryogenic Research Center, The University of Tokyo, 2-11-16 Yayoi, Bunkyo-ku, Tokyo, Japan\\        
}

\date{Received: date / Accepted: date}

\maketitle

\begin{abstract}
We report design details of the whole assembly of a compact and continuous nuclear demagnetization refrigerator (CNDR) with two PrNi$_5$ nuclear stages, which can keep temperature below 1~mK continuously, and test results of a new thermal contact method for the PrNi$_5$ stage using Zn soldering rather than Cd soldering.
By measuring a residual electrical resistance of a short test piece, the thermal contact resistivity between the PrNi$_5$ rod and an Ag wires thermal link was estimated as $1.8\pm0.1 \times 10^{-4} T^{-1}~\rm{Km^2W^{-1}}$.
Based on this value and 2D numerical and 1D analytical thermal simulations, the largest possible temperature gradient throughout the nuclear stage was calculated to be negligibly small ($\leq$ 2~\%) at 1~mK under a 10~nW heat leak, the expected cooling power of the CNDR.

\end{abstract}

\section{Introduction}
\label{sec: intro}
Recently, a sub-mK temperature environment is recognized as one of the frontiers in research fields of material science~\cite{Ref: Clark}, nanoelectronics~\cite{Ref: Palma}, and cryogenic particle detector~\cite{Ref: Shirron}.
The nuclear demagnetization refrigerator (NDR) with copper nuclear stage is a standard equipment to achieve such extremely low temperatures~\cite{Ref: Pobell}.
However, construction and operation of NDRs are technically demanding, which has been preventing non-experts from making use of the sub-mK environment.
In order to overcome the limitations, we recently proposed the concept of continuous refrigeration using two PrNi$_5$ nuclear demagnetization stages and a dilution refrigerator connected in series each other via two superconducting heat switches~\cite{Ref: Toda_1}.
The numerical simulations under realistic conditions showed that this new type of refrigerator, the continuous nuclear demagnetization refrigerator (CNDR), can keep the sample temperature at 0.8~mK with a cooling power of 10~nW~\cite{Ref: Toda_1}.
Developments of CNDR are now actively conducted~\cite{Ref: Schmoranzer_1,Ref: Takimoto,Ref: Schmoranzer_2}.

In this article, after showing an updated total design of our CNDR in Sec.~\ref{sec: design}, we report construction details of the PrNi$_5$ nuclear stage, one of major parts of the CNDR (Sec.~\ref{sec: nuclear stage}).
We tested a new soldering method with Zn for PrNi$_5$ and estimated a thermal contact resistivity of the Zn contact between a PrNi$_5$ rod and Ag wires from electrical resistance measurements (Sec.~\ref{sec: contact resistivity}).
Then, using known resistivities of all parts, a possible temperature gradient through the PrNi$_5$ nuclear stage was evaluated from numerical and analytical calculations on thermal models (Sec.~\ref{sec: temp. grad.}).

\section{Practical Design of the CNDR}
\label{sec: design}

\begin{figure*}
\centering
\includegraphics [width=0.75\textwidth]{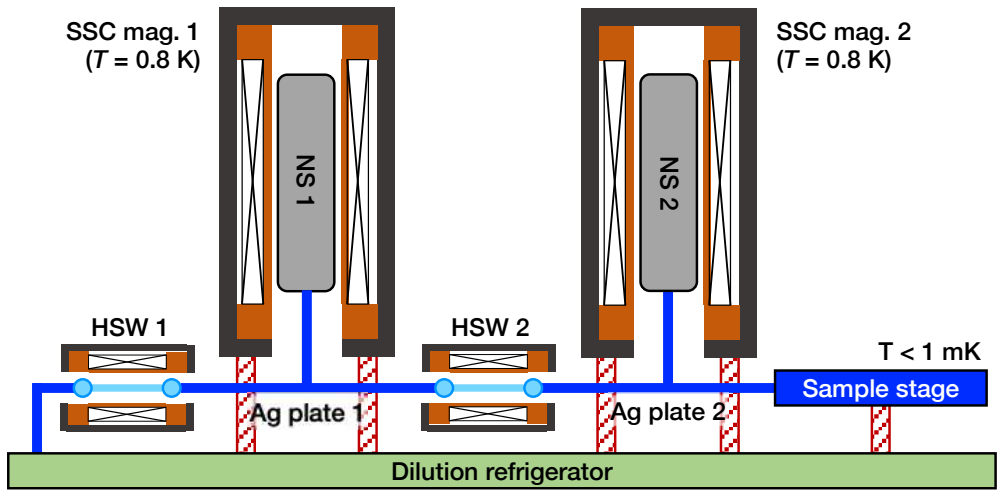}
\caption{Schematic figure of the CNDR. The two PrNi$_5$ nuclear stages with the two SSC magnets (SSC mag.1, 2) are connected in series through the two superconducting heat switches (HSW1, 2). All magnets are thermally anchored to the still of dilution refrigerator.}
\label{fig: CNDR schematic}
\end{figure*}

Figure~\ref{fig: CNDR schematic} shows a schematic diagram of our CNDR.
The CNDR consists of four major parts, (1) a standard dilution refrigerator which has a cooling power of at least 100~$\upmu$W at 100~mK, (2) two PrNi$_5$ nuclear stages, (3) two shielded superconducting (SSC) magnets~\cite{Ref: Takimoto} and (4) two superconducting Zn heat switches.
The PrNi$_5$ stages are connected in series between the sample stage and the mixing chamber of the dilution refrigerator through the two heat switches.
The CNDR can keep temperature below 1~mK continuously with a cooling power of 10~nW if the total thermal resistance among the four components is less than the value corresponding to a few hundreds n$\rm{\Omega}$ in the electrical resistance unit~\cite{Ref: Toda_1}. 

\begin{figure*}
\centering
\includegraphics[width=0.9\textwidth]{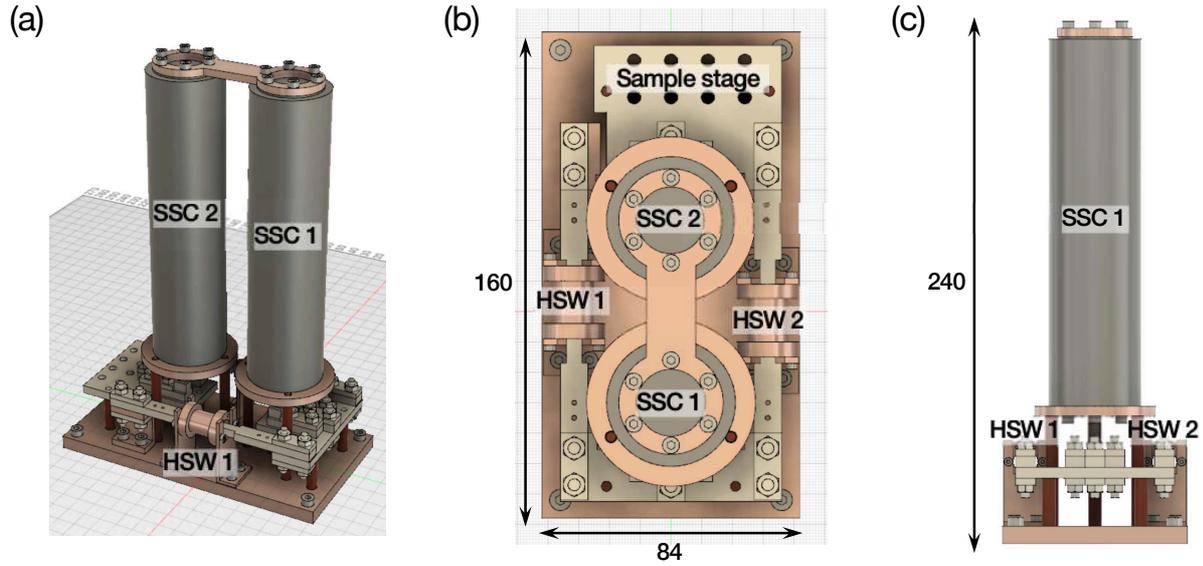}
\caption{Three-dimensional CAD model of the CNDR. (a) whole view, (b) top view, and (c) side view of the CNDR.
The unit of the dimensions denoted is millimeter.}
\label{fig: CNDR 3D}
\end{figure*}

Figure~\ref{fig: CNDR 3D} shows a three-dimensional CAD image of our latest version of CNDR.
The nuclear stages and other components are assembled on a Cu base plate which is connected directly to the mixing chamber.
The two SSC magnets are thermally anchored to the still of the dilution refrigerator and mechanically supported by Vespel SP-22 thermal insulation rods from the base plate.
Three of four Ag thermal links for the heat switches and six Ag thermal links for the nuclear stages are tightly connected to two Ag plates of 5~mm thick with M4 Si$_{0.15}$Ag$_{0.85}$ (Tokuriki Honten Co., Ltd.) screws so as to be demountable.
The cross section of each Ag thermal link is $9\times5$~mm$^2$.
The contact areas of these parts are gold-plated of 3--4~$\upmu$m thickness to reduce the contact thermal resistance~\cite{Ref: Okamoto}.
The maximum dimensions of the whole assembly of this CNDR are 160 mm in length, 84 mm in width and 240 mm in height, which are compact enough to be installable in most of dilution refrigerators.

\section{Zinc Soldering of PrNi$_5$ Nuclear Stage}
\label{sec: nuclear stage}
\begin{figure*}
\centering
\includegraphics[width=0.9\textwidth]{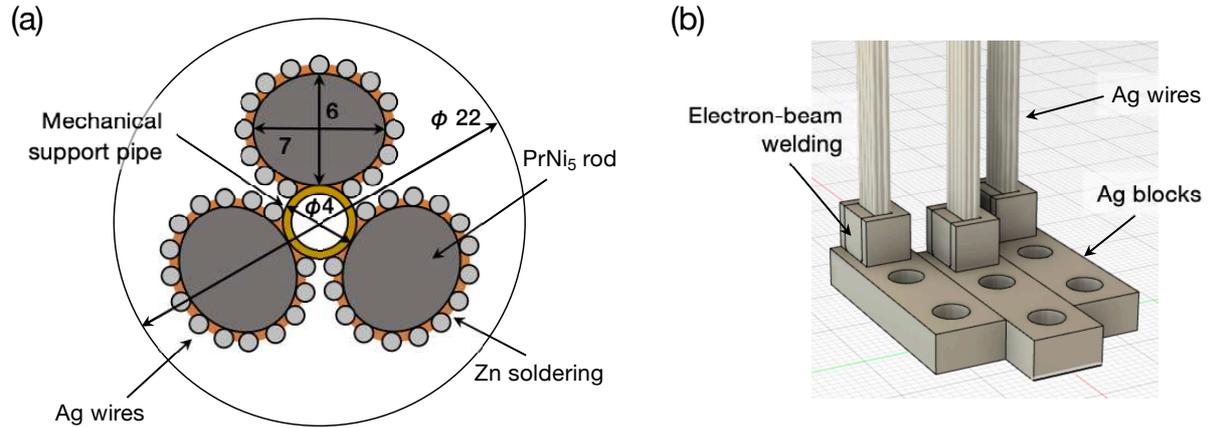}
\caption{(a) Cross section of the PrNi$_5$ nuclear stage. The bore diameter of the SSC magnet is 22~mm. (b) Far end of the nuclear stage. The Ag blocks e-beam welded with the Ag wires are attached to the Ag plate with silicon-silver screws.}
\label{fig: NS cross section}
\end{figure*}

It is important to achieve a better thermal contact between a nuclear coolant and a thermal link to minimize a temperature gradient between them under finite heat flows.
Figure~\ref{fig: NS cross section} (a) shows a cross section of the PrNi$_5$ nuclear stage of the CNDR.
Each stage consisting of three PrNi$_5$ rods of about 6~mm diameter and 120~mm long is soldered to the thermal link made of 15 Ag wires of 1~mm diameter with Zn as described in more detail later. 
It is noted that we cannot apply common press contact here because PrNi$_5$ is so brittle and can easily crack under stress.
The other ends of the Ag wires are electron-beam (EB) welded to the silver block (see Fig.~\ref{fig: NS cross section} (b)).
The depth and lateral dimensions of the welded part are 6~mm and 8$\times$6~mm$^2$, respectively.
After the EB welding, the assembled Ag thermal link was annealed at ${800^{\circ}}\mathrm{C}$ for 3 hours in an O$_2$ flow ($P \sim 0.1$~Pa).
The residual resistivity ratio (RRR) of the Ag wire was increased from 300 to 2,300 by this heat treatment.
From the four-wire measurement at 4.2~K, we determined a residual electrical resistance of the EB welded part to be $13.5 \pm 1.0$~n$\rm{\Omega}$.
This is lower than resistances of other parts throughout the nuclear stage (see later discussions). 
RRR of the PrNi$_5$ rod is 42.

There are three common soldering agents to attach the PrNi$_5$ rod to the thermal link, that are Cd~\cite{Ref: Mueller,Ref: Andres,Ref: Greywall,Ref: Parpia}, In~\cite{Ref: Wiegers} and Sn~\cite{Ref: Ishimoto_1}.
Cd has most commonly been used in the previous works because of its low superconducting critical field ($H_{\mathrm{c}} = 3.0$~mT).
Note that $H_{\rm{c}}$ limits the final field of demagnetization  cooling.
However, the problem of Cd is that it is toxic.
In has a ten times higher $H_{\rm{c}}$ than that of Cd and is mechanically rather weak.
Sn has also a high $H_{\rm c}$, and sometimes the cold brittleness due to structural transformation~\cite{Ref: Cohen} causes trouble.
In this work, as a substitute soldering agent for Cd, we tested to use Zn which has a low enough  $H_{\rm{c}}$ ($= 5.3$~mT) and low health damage.
A drawback of Zn is a little higher melting temperature (420~$^\circ$C) (see Table~\ref{tab: sold. agents}).

\begin{table*}
\caption{Various soldering agents for the PrNi$_5$/Ag contact.}
\label{tab: sold. agents}
\begin{center}
\begin{tabular}{lccccc}
\hline\noalign{\smallskip}
& \begin{tabular}{c} Melting temp.\\ $T_{\rm{m}}$~($^\circ$C)\end{tabular} & \begin{tabular}{c} Supercond.\\ $T_{\rm{c}}$~(K)\end{tabular} & \begin{tabular}{c} Supercond.\\ $H_{\rm{c}}$~(mT)\end{tabular} & Flux & Note \\
\noalign{\smallskip}\hline \hline\noalign{\smallskip}
Cd & 321 & 0.56 & 3.0 & ZnCl$_2$ or NH$_4$Cl & toxic \\
\noalign{\smallskip}
In & 157 & 3.40 & 29.3 & & high $H_{\rm{c}}$, soft \\
\noalign{\smallskip}
Sn & 232 & 3.75 & 30.9 & & high $H_{\rm{c}}$, brittle \\
\noalign{\smallskip}
Zn & 420 & 0.85 & 5.3 & ZnCl$_2$ + NH$_4$Cl & high $T_{\rm{m}}$ \\
\noalign{\smallskip}\hline
\end{tabular}
\end{center}
\end{table*}

\begin{figure*}
\centering
\includegraphics[width=0.9\textwidth]{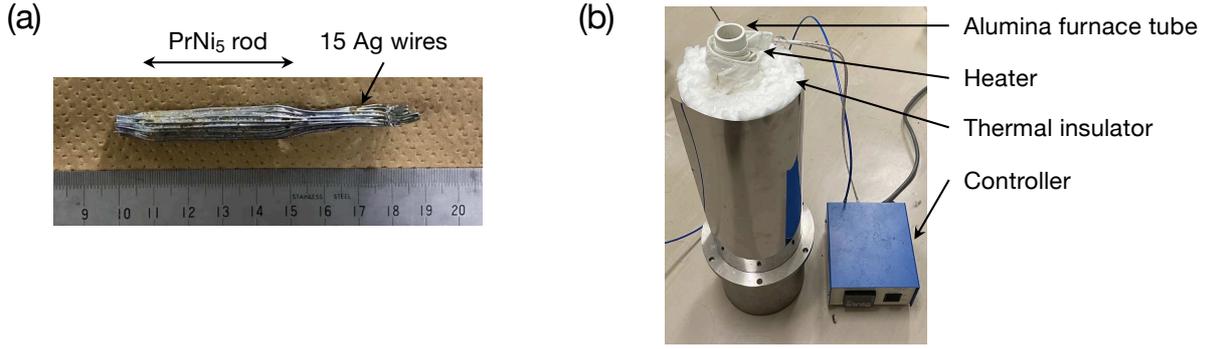}
\caption{(a) Short test piece for Zn soldering of the PrNi$_5$/Ag contact. (b) Home-made furnace used for the Zn soldering.}
\label{fig: soldering test}
\end{figure*}

We evaluated the performance of the Zn soldering for the PrNi$_5$ rod and Ag wires by making a short test piece shown in Fig.~\ref{fig: soldering test}(a).
Figure~\ref{fig: soldering test}(b) shows a home-made furnace used in this test where a heater wire is wound around a high-purity (99.6~\%) alumina furnace tube of 24~mm in inner diameter and 500~mm in height with a closed bottom.
The alumina tube, in which molten Zn is filled, are thermally insulated by a rock wool.
The tube temperature was kept around 480~$^\circ$C with a PID controller during the soldering.
The test piece was first immersed in a 10 wt\% water solution of NaOH kept at 80~$^\circ$C for 5 minutes for degreasing. After washing it with water, it was immersed in a 10 wt\% water solution of HCl kept at 30~$^\circ$C for 5 minutes for removing surface oxides. 
After washing it with water, it was then immersed in a flux, a 25 wt\% water solution of a mixture of ZnCl$_2$ (56 wt\%) and NH$_4$Cl (44 wt\%), kept at 60~$^\circ$C, and finally dipped in the molten Zn for 2 minutes for soldering.

\section{Measurement of the Zn Soldered PrNi$_5$/Ag Contact Resistance}
\label{sec: contact resistivity}

\begin{figure*}
\centering
\includegraphics[width=0.9\textwidth]{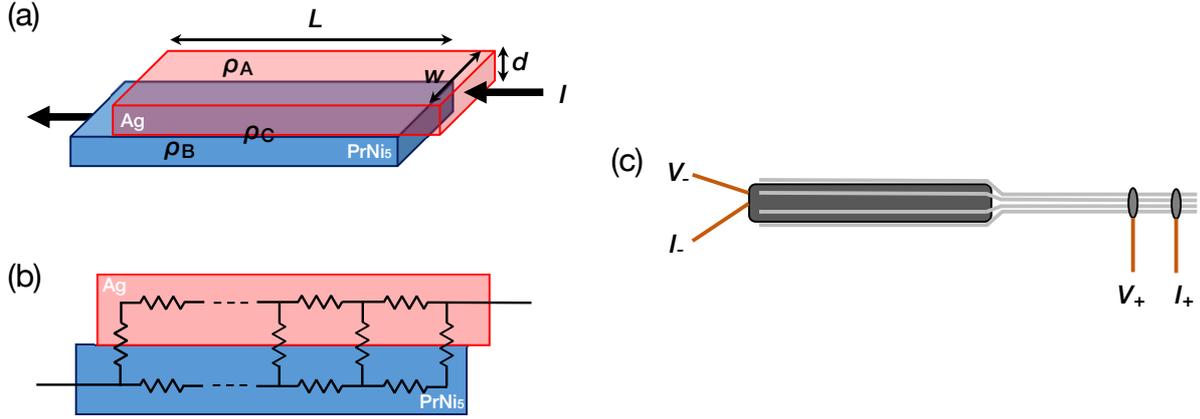}
\caption{(a) Schematic configuration for measurement of the thermal contact resistivity between the Ag wires and PrNi$_5$ rod. (b) Resistor ladder model for the configuration shown in (a). (c) Four-terminal setup for the $R^{\rm e}_{\rm{tot}}$ measurement to estimate the contact resistivity in the Zn soldered PrNi$_5$ nuclear stage.}
\label{fig: contact model}
\end{figure*} 

The thermal contact resistivity of the Zn-soldered Ag and PrNi$_5$ contact was evaluated from a measurement of the total resistance of the test piece in the configuration shown in Fig.~\ref{fig: contact model}(a).
The red and blue pieces here correspond to the Ag wires (A) and the PrNi$_5$ rod (B), respectively.
The contact area between A and B is denoted as C.
The total resistance $R_{\rm{tot}}$ of this configuration is analytically given as the following equation~\cite{Ref: Ishimoto_2} based on the ladder model shown in Fig.~\ref{fig: contact model}(b):
\begin{multline}
\label{eq: Rtot}
R_{\rm{tot}} = \frac{r_{\rm{A}} r_{\rm{B}}}{r_{\rm{A}} + r_{\rm{B}}}l + \frac{r_{\rm{A}}^2 + r_{\rm{B}}^2}{r_{\rm{A}} + r_{\rm{B}}}l_0 \coth \left( \frac{l}{l_0} \right) \\
+ \frac{2 r_{\rm{A}} r_{\rm{B}}}{r_{\rm{A}} + r_{\rm{B}}}l_0 \left[ \sinh \left( \frac{l}{l_0} \right) \right]^{-1},
\end{multline}
where
\begin{equation}
r_{\rm{A}} = \frac{\rho_{\rm{A}}}{w_{\rm{A}} d_{\rm{A}}}, \ r_{\rm{B}} = \frac{\rho_{\rm{B}}}{w_{\rm{B}} d_{\rm{B}}}, {\rm{and}} \ l_0 = \sqrt{\frac{\rho_{\rm{C}}}{(r_{\rm{A}} + r_{\rm{B}}) w_{\rm{C}}}}.
\end{equation}
Here, $\rho_{\rm{A}}$ and $\rho_{\rm{B}}$ are volume resistivities of A and B, respectively, and $\rho_{\rm{C}}$ is a contact resistivity. $l$ and $w$ are the length and the width of the contact area, respectively, and $d$ is the thickness of each piece.
Since the ladder model is applicable to both thermal (t) and electrical (e) flow problems, superscripts t or e will be put on variables when we want to specify either t- or e-resistance in the following.

We measured $R^{\rm e}_{\rm{tot}}$ of the test piece at 4.2~K in liquid $^4$He by using the four-terminal method shown in Fig.~\ref{fig: contact model}(c).
$\rho^{\rm e}_{\rm{A}}$ and $\rho^{\rm e}_{\rm{B}}$ were determined from independent measurements for each Ag and PrNi$_5$ pieces at 4.2~K.
By substituting these values to eq.(\ref{eq: Rtot}), we have $\rho^{\rm e}_{\rm{C}} = 4.3\pm0.2~\rm{p\Omega m^2}$.
Since conduction electrons are carriers of both electrical and thermal flows in pure metals and metallic compounds at millikelvin temperatures, one can convert $\rho^{\rm e}$ to $\rho^{\rm t}$ assuming the Wiedemann-Franz law with the Lorenz number $L$ ($= 2.45 \times 10^{-8}$~W$\rm{\Omega}$K$^{-2}$).
The fairly good applicability of this law to PrNi$_5$ is verified down to 100~mK~\cite{Ref: Meijer}.
Then, we have the thermal contact resistivity as $\rho^{\rm t}_{\rm{C}} = (1.8\pm0.1) \times 10^{-4} T^{-1}~\rm{Km^2W^{-1}}$.

The obtained electrical resistivity $\rho^{\rm e}_{\rm{C}}$ of the Zn soldered PrNi$_5$/Ag contact, which gives a total contact resistance of 4~n$\rm{\Omega}$ for each PrNi$_5$ rod, is sufficiently low for our purpose.
This can easily be understood if compared with the total electrical resistance of 140~n$\rm{\Omega}$ throughout the whole PrNi$_5$ stage in Greywall's single-stage nuclear demagnetization refrigerator~\cite{Ref: Greywall}.
In the next section, we will confirm this more quantitatively.

\section{Simulations for Temperature Gradient in the Nuclear Stage}
\label{sec: temp. grad.}

\begin{figure*}
\centering
\includegraphics[width=0.9\textwidth]{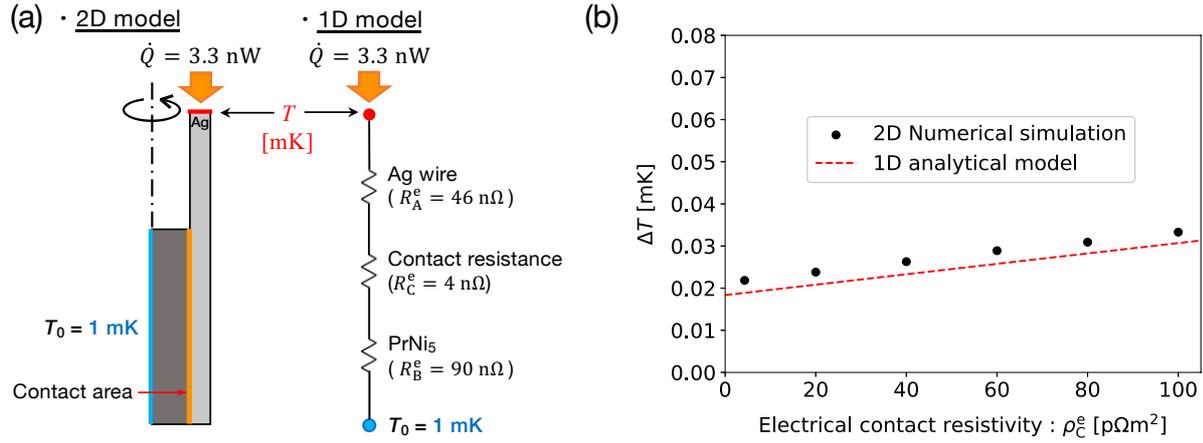}
\caption{(a) Two types of thermal models to estimate the largest possible temperature gradient ($\Delta T$) thoughout the PrNi$_5$ nuclear stage. (left) 2D model for numerical simulation by the finite element method. (right) 1D analytical model. (b) $\Delta T$ as a function of the Zn-soldering contact resistivity ($\rho^{\rm e}_{\rm{C}}$) estimated from the 2D (dot) and 1D (dashed line) models.}
\label{fig: temp. grad.}
\end{figure*}

Using the $\rho^{\rm t}_{\rm{A}}$, $\rho^{\rm t}_{\rm{B}}$ and $\rho^{\rm t}_{\rm{C}}$ values obtained in the previous section, one can evaluate a temperature difference $\Delta T$ ($> 0$) throughout the nuclear stage under a given heat flow.
In order to estimate the largest possible $\Delta T$, we made numerical simulations based on the finite element method under a constrain that the temperature ($T_0$) along the central ($z$) axis of the PrNi$_5$ rod is always kept at 1~mK as shown in Fig.~\ref{fig: temp. grad.}(a) left.
The constrain is equivalent to neglecting cooling powers associated with slow demagnetization of nuclear spins in off-axial segments within the PrNi$_5$ rod. 
This intentionally increases a radial ($r$) temperature gradient than the actual situation.
For simplicity, we also applied axisymmetry about the central axis to the model.
This means that no azimuthal component of heat flow is considered.
A constant heat flow of $\dot{Q} = 3.3$~nW, which is one third of the expected total cooling power of our CNDR, was introduced from the top end of the Ag wires (see Fig.~\ref{fig: temp. grad.}(a) left).
Thus the temperature at that end is $T_0 + \Delta T$.
The simulations were carried out using an open source software (FEMM: Finite Element Method Magnetics by David Meeker).
In the simulation, a $z$--$r$ (two dimensional: 2D) plane of the nuclear stage is divided to 16,000 elements.

Results of the simulation for six different $\rho^{\rm e}_{\rm C}$ ($= 4.3, 20, 40, 60, 80, 100$~p${\rm \Omega m^2}$) are shown as the dots in Fig.~\ref{fig: temp. grad.}(b).
For the $\rho^{\rm e}_{\rm{C}}$ value ($= 4.3\pm0.2~\rm{p\Omega m^2}$) we obtained with the Zn soldering, $\Delta T$ is expected to be less than 20~$\upmu$K at $T_0 = 1$~mK, which is negligibly small.
$\Delta T/T_0$ is not sensitive to $\rho^{\rm e}_{\rm{C}}$ and will not exceed 10\% unless $\rho^{\rm e}_{\rm{C}}$ becomes larger by two orders of magnitude. 

The dashed line in Fig.~\ref{fig: temp. grad.}(b) is an expected behavior from the one-dimensional (1D) model considering serially connected three thermal resistances which are the resistance of the Ag wires ($R^{\rm t}_{\rm{A}}$) above the stage, that of the PrNi$_5$ ($R^{\rm t}_{\rm{B}}$) and the contact resistance ($R^{\rm t}_{\rm{C}}$) between them (see Fig.~\ref{fig: temp. grad.}(a)  right).
$R^{\rm t}_{\rm{B}}$ and $R^{\rm t}_{\rm{C}}$ are calculated from $\rho^{\rm t}_{\rm{B}}$ and $\rho^{\rm t}_{\rm{C}}$ neglecting the resistance of the Ag wires over the length of the PrNi$_5$ rod and considering only the $r$ component of heat flow.
This simplification resembles to the constraint of fixed $T_0$ (= 1~mK) along the $z$ axis in the 2D numerical simulation described above.
The analytical solution of the 1D model is expressed as:

\begin{equation}
\label{eq: temp. grad.}
\Delta T = \sqrt{T_0^2 + \frac{\dot{Q}}{\pi L} \left( A \cdot \rho^{\rm e}_{\rm{A}} + B \cdot \rho^{\rm e}_{\rm{B}} + C \cdot \rho^{\rm e}_{\rm{C}} \right)} - T_0,
\end{equation}
where $A (= 4.27 \times 10^4$~m$^{-1})$, $B (= 28.3$~m$^{-1})$, and $C (= 5.85 \times 10^3$~m$^{-2})$ are geometrical factors.
As can be seen in Fig.~\ref{fig: temp. grad.}(b), the 1D analytical model gives slightly smaller $\Delta T$ values than the 2D numerical simulation, because the temperature variation along the $z$ direction is neglected.
However, the difference (8--16~\%) is small and the 1D model is much simpler to calculate, so it is a more convenient estimator of $\Delta T$.

\section{Conclusion}
\label{sec: conclusion}
We described design details of the whole assembly of the continuous nuclear demagnetization refrigerator (CNDR) which is so compact that it can be installed on the mixing chamber of an existing dilution refrigerator. 
We focused on the design of the PrNi$_5$ nuclear stage, a central part of the CNDR, in this article.
As an alternative to the widely-used Cd soldering thermal contact for PrNi$_5$, the Zn soldering was proposed and tested.
From the residual electrical resistance measurements of a short test piece, the thermal contact resistivity between the PrNi$_5$ rod and the Ag wires was estimated as $1.8\pm0.1 \times 10^{-4} T^{-1}~\rm{Km^2W^{-1}}$ assuming the Wiedemann-Franz law.
This is favorably compared with the previously reported contact resistivities for other metals and soldering agents.
Using known resistivities of all major parts of the CNDR, we evaluated the largest possible temperature gradient $\Delta T / T$ throughout the nuclear stage from the 2D numerical and 1D analytical calculations.
The calculations show a negligibly small $\Delta T / T$ ($\leq 2$\%) at 1~mK under a 10~nW heat leak, which is an expected cooling power of the CNDR.
All parts of the CNDR are now being assembled anticipating the first cooling test.

\begin{acknowledgements}
We thank the machine shop of the School of Science, the University of Tokyo for machining most of the parts of CNDR.
ST was supported by Japan Society for the Promotion of Science through Program for Leading Graduate Schools (MERIT).

\end{acknowledgements}

%
%



\end{document}